\documentclass[aip,apl,preprint]{revtex4-1}
\usepackage{graphicx}
\usepackage{natbib}
\usepackage{amsmath}
\usepackage{amssymb}
\usepackage{bm}
\usepackage{color}
\def\etal{$\it{et~al.}$}

\begin{document}

\title{Point-contact Andreev reflection spectroscopy of candidate topological superconductor Cu$_{0.25}$Bi$_2$Se$_3$}

\author{X. Chen}
\email{xcchen@gatech.edu}
\affiliation{School of Physics, Georgia Institute of Technology, Atlanta, GA 30332, USA}

\author{C. Huan}
\affiliation{School of Physics, Georgia Institute of Technology, Atlanta, GA 30332, USA}

\author{Y. S. Hor}
\affiliation{Department of Physics, Missouri University of Science and Technology, Rolla, MO 65409, USA}

\author{C. A. R. S\'a de Melo}
\affiliation{School of Physics, Georgia Institute of Technology, Atlanta, GA 30332, USA}

\author{Z. Jiang}
\email{zhigang.jiang@physics.gatech.edu}
\affiliation{School of Physics, Georgia Institute of Technology, Atlanta, GA 30332, USA}


\begin{abstract}
We perform a point-contact Andreev reflection spectroscopic study of the topological superconducting material, Cu$_{0.25}$Bi$_2$Se$_3$, in the ballistic regime using a normal-metal gold tip. We observe distinct point-contact spectra on the superconducting and non-superconducting regions of the crystal surface: the former shows a marked zero-bias conductance peak, indicative of unconventional superconductivity, while the latter exhibits a pseudogap-like feature. In both cases the measured differential conductance spectra exhibit a large linear background, preventing direct quantitative comparison with theory. We attribute this background to inelastic scattering at the tip-sample interface, and compare the background-subtracted spectra with a single-band \textit{p}-wave model.
\end{abstract}

\maketitle

Topological superconductors (TSCs) have drawn substantial attention lately, owing to the recent research interest in topological order and its possible application to quantum computing \cite{TI_review}. In analogy to topological insulators, TSCs possess gapless states on the surface that are protected by a topological invariant, while the bulk of the material is superconducting (consisting of a fully opened energy gap). It has been theoretically predicted that the pairing symmetries of TSCs are unconventional, and robust zero-energy Majorana bound states may appear on the surface \cite{schnyder_08,kitaev_09,qi_09,pwave_CBS,fu_10,hao_11,majorana}. Correlated (Bogoliubov) quasiparticles in such states may exhibit non-Abelian statistics \cite{moore_91}, forming the basis for realizing the intriguing proposals of fault-tolerant quantum computing \cite{kitaev_03,quantum_computing}. Creation, detection and manipulation of non-Abelian quasiparticles in strongly correlated systems have been a frontier of fundamental research in the past ten years. The emergent TSCs have added to this endeavor an intriguing possibility: the existence of Majorana fermions.

The experimental investigation of time-reversal-invariant TSCs has just started, triggered by the discovery of superconductivity in the topological material, Cu$_{x}$Bi$_2$Se$_3$ \cite{hor_sample,wray_10,ando_sample,kriener_11}, which has a transition temperature $T_c$$\sim$3.8 K just below the liquid $^4$He temperature of 4.2 K. The spin-orbit coupling between the motion of conduction electrons and their spin is expected to be substantially large in this material, leading to a triplet component of the order parameter, which is largely responsible for its unconventional superconductivity and other exotic physical properties. A dipper understanding of the superconductivity of Cu$_{x}$Bi$_2$Se$_3$ can shed light on the solid-state realization of Majorana fermions and provide valuable guidance for the current search of other TSCs \cite{sasaki_new}.

Point-contact Andreev reflection spectroscopy has proved to be a useful tool in exploring superconducting materials, particularly unconventional superconductors \cite{deutscher_05}. It provides both the energy- and the momentum-resolved spectroscopic information at the interface of two materials (usually one is known and the other is unknown). It has been widely used in determining the pairing symmetry of $p$-wave and $d$-wave superconductors, and quantitative agreement with theory has been achieved \cite{laube_00,liu_01,yakovenko_02,dwave_PCAR}. Recently, this technique has also been employed to study the superconductivity of Cu$_x$Bi$_2$Se$_3$ \cite{Sasaki,Kirzhner}, but the results were inconclusive due to the complexity of the material and the unusually large background in the obtained point-contact spectra. Here, we present a point-by-point comparison of the Andreev reflection spectra measured at different locations on the surface of a Cu$_{0.25}$Bi$_2$Se$_3$ single crystal. We find that the spectrum background is approximately linear at high bias voltages, and the slope is independent of temperature but varies from place to place. These observations suggest that the background may stem from inelastic scattering at the interface, and is therefore unrelated to the superconducting properties of the material.

The Cu$_{0.25}$Bi$_2$Se$_3$ single crystal studied in this work was prepared by melting stoichiometric amounts of copper (99.99$\%$), bismuth (99.999$\%$) and selenium (99.999$\%$) in a sealed quartz tube, followed by slow cooling from 850 $^{\circ}$C to 620 $^{\circ}$C and finally quenching in cold water. Transport characterization \cite{luli_CBS} reveals that the superconducting transition starts from $\sim$3.3 K and the zero resistance state is reached at $\sim$1.2 K. The nominal superconducting fraction is found to be $\sim$35$\%$, consistent with the phase diagram of Cu$_x$Bi$_2$Se$_3$ reported by Kriener \textit{et al.} \cite{kriener_11}

\begin{figure}
\includegraphics[width=10cm]{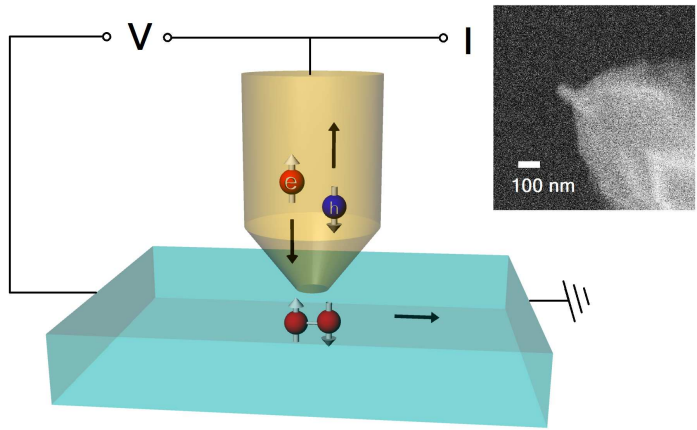}
\caption{\label{fig1}(color online) Schematics of the experimental set-up and the Andreev reflection process at the normal-metal/superconductor interface. An electron (red) enters the  superconductor by forming a Cooper pair with another electron of opposite spin (white arrow for spin), while leaving a hole (blue) reflected back from the interface to the normal metal. Inset: SEM image of a gold tip.}
\end{figure}
In Fig. \ref{fig1}, we show a highly simplified schematic version of our experimental set-up, where a normal-metal gold tip is used to approach the surface of Cu$_{0.25}$Bi$_2$Se$_3$ single crystal. In order to create a ``point'' contact, we etch electrochemically a gold wire (99.999$\%$ pure and 0.25 mm in diameter) in a HCl(37$\%$):ethanol (1:1) solution, following the procedure described in Ref. [\onlinecite{gold_tip}]. The inset in Fig. \ref{fig1} shows a scanning electron microscope (SEM) image of a gold tip, where the tip end is less than 100 nm in diameter. The Cu$_{0.25}$Bi$_2$Se$_3$ sample was mounted on a stack of linear nanopositioners (attocube systems), which allowed for accurate control of the tip-sample distance as well as measurements at multiple surface locations. The set-up was thermally anchored to a $^3$He cryostat equipped with a 14 T superconducting magnet. In our experiment, the point-contact spectra were taken on the freshly cleaved crystal surface to minimize possible oxidation and copper segregation. The system implements feedback from the control program of nanopositioners during the tip approach, to ensure a small footprint of the point contact and ballistic transport through the interface. The typical contact resistance at low temperatures is $R_c$$\approx$10 $\Omega$, which corresponds to an effective contact diameter $d$$\approx$12 nm estimated from the Sharvin resistance formula, $R_c=\frac{16\rho l }{3\pi d^{2}}$, where $\rho$ and $l$ are the resistivity and the mean free path of the gold tip, respectively. In our experiment, the condition $d$$<<$$l$ is satisfied, and thus ballistic transport can be assumed, since $l$$\approx$38 nm at room temperature and is expected to be much longer at low temperatures. A standard dc $+$ ac lock-in technique was used to measure the point-contact spectra (\textit{i.e.}, $dI/dV$ vs. bias voltage) of the tip-sample interface. The ac current applied was 100 nA.

\begin{figure}[t]
\includegraphics[width=10cm]{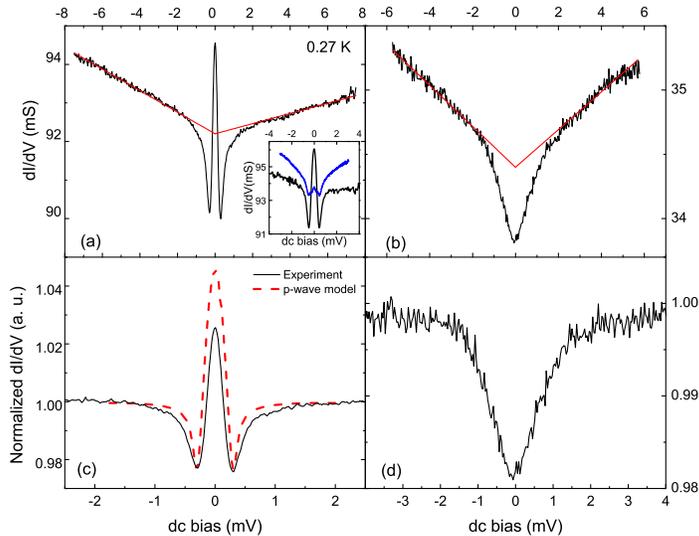}
\caption{\label{fig2}(color online) Typical $dI/dV$ spectra of the point contact on (a) superconducting and (b) non-superconducting regions of the Cu$_{0.25}$Bi$_2$Se$_3$ surface. Inelastic tunneling through the interfacial barrier gives rise to the asymmetric, linear background (red lines) in the spectra. Inset: The background exhibits different slopes at different measurement locations, presumably due to different barrier strengths. Panels (c) and (d): Normalized $dI/dV$ spectra after removal of the background in (a) and (b). The dashed red line in (c) is a comparison with the single-band $p$-wave model described in the text.}
\end{figure}
Figures \ref{fig2}(a) and \ref{fig2}(b) present the typical point-contact spectra measured at 0.27 K. For certain surface locations, we observed a pronounced zero-bias conductance peak (ZBCP) in the spectra, indicative of unconventional superconductivity. For the other locations, however, only a broad dip is seen at low bias voltages. Since our measurement was performed in the ballistic regime, the observed ZBCP was not caused by heating effects, which may give rise to spurious ZBCP in the diffusive/thermal regime \cite{heating}. In addition, heating effects are not responsible for the asymmetric, linear background prominent at high bias voltages, because one would expect a decrease in conductance due to heating rather than an increase. We find that this linear background occurs in all the spectra that we have measured; it is nearly independent of temperature, and it varies from place to place on the Cu$_{0.25}$Bi$_2$Se$_3$ surface. Such a background was also observed in previous studies of Cu$_{x}$Bi$_2$Se$_3$ \cite{Kirzhner,Sasaki}, but its origin was not explained. Here, we attribute it to inelastic tunneling at the tip-sample interface, following the model of Kirtley \etal \cite{kirtley1,linear_bg,kirtley2} Since the Cu$_x$Bi$_2$Se$_3$ sample is very sensitive to its environment, it is not surprising that oxidation barriers can form on the surface. A broad, flat distribution of energy-loss modes in the barrier could lead to the asymmetric, linear background appearing in the $dI/dV$ spectra
\begin{equation}
\label{background}
(dI/dV)_{inelastic}\approx\int_0^{eV}F(E)dE.
\end{equation}
Here, $eV$ is the electron energy and $F(E)$$\approx$$constant$ is the spectral distribution of inelastic scattering modes. In Figs. \ref{fig2}(c) and \ref{fig2}(d), we plot the point-contact spectra after removal of the background. The spectra are normalized with respect to the $dI/dV$ values at high bias voltages (normal state) in order to compare with theory.

\begin{figure}[t]
\includegraphics[width=8cm]{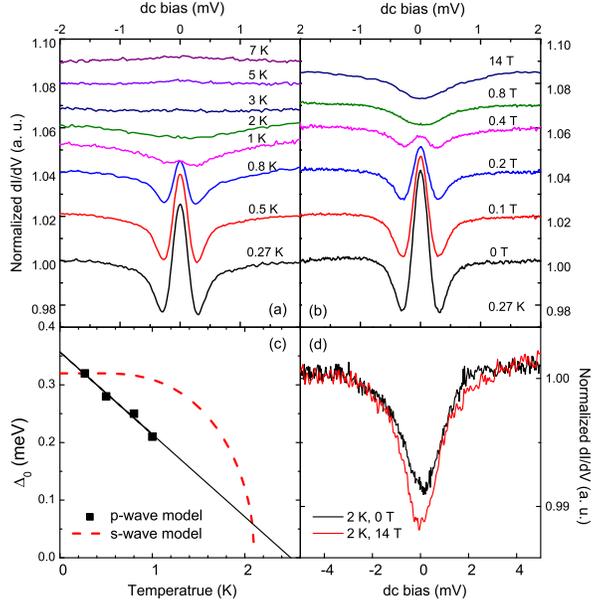}
\caption{\label{fig3}(color online) (a) Temperature and (b) magnetic field dependence of the point-contact spectra on the superconducting region of the sample. The spectra are normalized with respect to the $dI/dV$ values at 4 mV, and shifted vertically for clarity. The magnetic field applied is perpendicular to the $ab$ plane. (c) Temperature dependence of the superconducting gap (filled squares) extracted from the single-band $p$-wave model described in the text, and compared with the expected values from the \textit{s}-wave BCS theory (dashed red line). (d) $dI/dV$ spectra obtained on the non-superconducting region of the sample at 0 T (black) and 14 T (red). An asymmetric, linear background has been subtracted from all the spectra.}
\end{figure}
Figures \ref{fig3}(a) and \ref{fig3}(b) show the temperature and magnetic field dependence of the ZBCP after background subtraction. We find that the ZBCP only occurs at temperatures below the $T_c$ of Cu$_{0.25}$Bi$_2$Se$_3$ and in a magnetic field below the upper critical field ($B_{c2}$=1.7 T for Cu$_{0.29}$Bi$_2$Se$_3$) \cite{ando_sample}, while the other type of spectral lineshape, as shown in Figs. \ref{fig2}(b) and \ref{fig2}(d), is only weakly dependent on the temperature and the magnetic field (Fig. \ref{fig3}(d)). Therefore, we conclude that Figs. \ref{fig2}(a) and \ref{fig2}(b) represent the point-contact spectra on the superconducting and non-superconducting regions of the sample, respectively. In addition, the spectra shown in Figs. \ref{fig2}(b) and \ref{fig2}(d) exhibit a pseudogap-like feature, \textit{i.e.}, the reduction of the density of states at low bias voltages, which appears to be enhanced by the magnetic field (Fig. \ref{fig3}(d)). Similar behavior has also been observed by Sasaki \textit{et al.} \cite{Sasaki} on the superconducting region at $T$$>$$T_c$, suggesting that the \textit{non-superconducting} region referred herein may become superconducting at lower temperatures. We note that the observed magnetic field dependence is unexpected, markedly different from the pseudogap behavior in high-temperature superconductors. Further experimental and theoretical work is needed to understand this observation.

The presence of the ZBCP and the absence of coherence peaks near the superconducting gap edge in Figs. \ref{fig2}(a) and \ref{fig2}(c) certainly cannot be explained by the tunneling theory of conventional (\textit{s}-wave) superconductors \cite{btk}. Alternative explanations such as reflectionless tunneling \cite{beenakker_91,vanwees_92} or magnetic and Kondo scattering \cite{appelbaum,shen_68} can also be ruled out, following the analysis by Sasaki \textit{et al.} \cite{Sasaki} Moreover, one can show that proximity-induced supercurrent \cite{kastalsky_91} is not relevant to our observations, since the induced supercurrent is expected to vanish in a very low magnetic field ($\sim$0.02 T), inconsistent with the magnetic field dependence of Fig. \ref{fig3}(b). Therefore, we conclude further that the observed ZBCP is most likely due to the surface midgap Andreev bound states (ABS) caused by the unconventional superconductivity of the material \cite{ABS,hu_94,tanaka_95}.

Recently, advanced theories have been developed to describe the pairing symmetry of Cu$_x$Bi$_2$Se$_3$ and the nature of its surface ABS \cite{fu_10,hao_11,Sasaki,fu_12}. Since the exact form of the interaction is not known in this material, theoretically one could assume pairing in all bands is possible and consider all possible types of pairing symmetries. Experimentally, however, we find that no direct evidence of multiple bands is observed in point-contact spectra, which suggests that an effective \textit{single-band} model might be sufficient to describe Cu$_x$Bi$_2$Se$_3$. This situation is similar to encountered high-temperature superconductors, where a somewhat complicated three-band theory is replaced by an effective single-band model \cite{zhang_88} to explain experimental results. Therefore, it is enlightening to compare our data with existing single-band tunneling theory for triplet \textit{p}-wave pairing. Towards this end, we apply the model of Yamashiro \etal \cite{pwave_btk} to calculate the $dI/dV$ spectra in the clean limit
\begin{equation}
\label{dIdV}
dI/dV=D_0\int_{-\infty}^{+\infty}dE\ \sigma(E) \left[\frac{-\partial f(E+eV)}{\partial E}\right].
\end{equation}
Here, $D_0$ is the transmission probability that rescales the calculated spectra with respect to the data \cite{laube_00}, $\sigma(E)$ is the normalized tunneling conductance as described in Ref. [\onlinecite{pwave_btk}], and $f(E)$ is the Fermi function. For the sake of simplicity, we ignore the hexagonal symmetry of the crystal in our calculation and choose a simple pairing potential, $\Delta_{\uparrow\uparrow}=\Delta_0$sin$\theta e^{i\phi}$ and $\Delta_{\uparrow\downarrow}= \Delta_{\downarrow\uparrow}=\Delta_{\downarrow\downarrow}=0$, where $\uparrow$ and $\downarrow$ denote spin indices, $\Delta_0$ is the amplitude of the order parameter in the bulk, and $\theta$ and $\phi$ are the polar angle and azimuthal angle, respectively. As discussed in Ref. [\onlinecite{pwave_btk}], the appearance of the ZBCP depends on the tunneling direction; it only appears in the in-plane tunneling spectra (within the $ab$ plane), in which the tunneling occurs from the tip to its neighboring sample surface areas (possibly through step edges). Figure \ref{fig2}(c) shows the comparison of our data with the calculated in-plane tunneling spectra: the ZBCP and the positions of the $dI/dV$ minima are well captured using $D_0$=0.071, $\Delta_0$=0.32 meV, and $Z$=10, where $Z$ characterizes the interfacial barrier strength \cite{btk}. More comprehensive models considering lifetime broadenings and/or nodes in the order parameter may be able to fit our data quantitatively, but this is beyond the scope of this work.

Finally, it is worth noting that the superconducting gap extracted from the above \textit{p}-wave model is always slightly larger than the corresponding energies of the $dI/dV$ minima in the point-contact spectra. In Fig. \ref{fig3}(c), we plot the temperature dependence of $\Delta_0(T)$, which appears linear in temperature, and therefore cannot be fitted to the standard Bardeen-Cooper-Schrieffer (BCS) theory for conventional (\textit{s}-wave) superconductors.

In summary, we have shown that in the ballistic limit, the point-contact spectra of Cu$_{0.25}$Bi$_2$Se$_3$ are similar to that observed by Sasaki \textit{et al.} \cite{Sasaki} who used a ``soft'' point-contact technique. In this work, we have argued that the observed asymmetric, linear background in the $dI/dV$ spectra is due to inelastic scattering through the interfacial barrier, and that a simple single-band \textit{p}-wave model could capture the main features of the $dI/dV$ spectra, in particular, the ZBCP.

We would like to thank D. B. Torrance and P. N. First for assistance in preparing high-quality gold tips. We acknowledge support from the American Chemical Society Petroleum Research Fund and the Army Research Office (W911NF-09-1-0220). TSC material synthesis is supported by MRC, Missouri S\&T.

\end{document}